\documentclass[12pt]{article}
\usepackage{geometry}             
\usepackage{graphicx}
\usepackage{amssymb}
\usepackage{amsmath}
\usepackage{epstopdf}
\usepackage{subfigure}
\usepackage{cite}

\usepackage[hyperindex=true,
          pdfstartview=FitH,
          bookmarksnumbered=true,
          bookmarksopen=true,
          citecolor=blue,
          linkcolor=blue,
          colorlinks=true,
          pdfborder=001,
          unicode]{hyperref}

\begin{document}

\title{Dyonic Born-Infeld black hole in four-dimensional Horndeski gravity}
\author{Kun Meng\thanks{email: {\it kunmeng@wfu.edu.cn}}, Lianzhen Cao, Jiaqiang Zhao, \\
 Tao Zhou, Fuyong Qin\thanks{email: {\it 811709892@wfu.edu.cn}},  Meihua Deng\thanks{email: {\it mhdeng@wfu.edu.cn}}\\
School of Physics and Photoelectric Engineering, Weifang University, \\
 Weifang 261061, China\\
}
\date{}                             
\maketitle

\begin{abstract}
The action of four-dimensional Horndeski gravity coupled to Born-Infeld electromagnetic fields is given via the Kaluza-Klein process. Dyonic black hole solution of the theory is constructed.
The metric is devoid of singularity at the origin independent of the parameter selections, this property is different from the one of Einstein-Born-Infeld black holes. Thermodynamics of the black hole is studied, thermodynamic quantities are calculated and the first law is checked to be satisfied. Thermodynamic phase transitions of the black holes are studied in extended phase space.
\end{abstract}

\section{Introduction}
Lovelock theorem states that, in four dimensions the tensors that satisfy divergence free, symmetric, and concomitant of the metric tensor and its derivatives  are no more than the metric tensor and the Einstein tensor\cite{Lovelock:1971yv}. That is to say Einstein's general relativity (GR) is the unique proper theory of gravity in four dimensions. Recently, in order to bypass Lovelock theorem a proposal has been made by adding Gauss-Bonnet term to  GR\cite{Glavan:2019inb}. As we know, the Gauss-Bonnet contribution is a topological invariant in four dimensions according to Gauss-Bonnet theorem, and does not affect the field equations of the theory. The authors of Ref.\cite{Glavan:2019inb} first make the replacement $\alpha\rightarrow\frac{\alpha}{D-4}$ to cancel the factor $D-4$ in the Gauss-Bonnet term contribution to field equations, and then take the $D\rightarrow4$ limit. The resulting four-dimensional(4D) Einstein-Gauss-Bonnet(EGB) gravity  exhibits novel properties not found in GR. Exact solutions of the gravity theory are studied in\cite{2004.14468,2003.07769,2006.07913,2003.14275,2004.07934,2003.02523,2003.13068,2004.03141}.

However,  Wald entropy of the black hole in the 4D EGB gravity
is divergent manifestly in the $D\rightarrow4$ limit. The standard thermodynamic relation  between free energy and entropy $I=-T^{-1}F=S-T^{-1}M$ also implies divergence in the on-shell Euclidean action\cite{2004.09214}, thus the action cannot account for the Euclidean path-integral for the topologically nontrivial solutions.
Meanwhile, the Gauss-Bonnet contribution to field equations $\mathcal{H}_{\mu\nu}$ can be decomposed into two parts
$\mathcal{H}_{\mu\nu}=-2(\mathcal{L}_{\mu\nu}+\mathcal{Z}_{\mu\nu})$,
where the $\mathcal{Z}_{\mu\nu}$ part is proportional to $D-4$, it is regular after the  rescaling  $\alpha\rightarrow\frac{\alpha}{D-4}$ and taking the $D\rightarrow4$ limit. While the tensor $\mathcal{L}_{\mu\nu}$, which can be expressed in terms of Weyl tensor
$
\mathcal{L}_{\mu\nu}=C_{\mu\alpha\beta\sigma}C_\nu^{\;\;\alpha\beta\sigma}-\frac{1}{4}g_{\mu\nu}C_{\alpha\beta\rho\sigma}C^{\alpha\beta\rho\sigma},
$
vanishes identically in $D\leq4$,  thus $\frac{\mathcal{L}_{\mu\nu}}{D-4}$ is undefined\cite{2004.08362,2009.13508,2004.12998}.

In order to add the Gauss-Bonnet contribution regularly and nontrivially to Einstein gravity in four dimensions, one way is to take the Kaluza-Klein reduction of the EGB theory in higher dimensions. The authors of Ref.\cite{2003.11552} compactify  D-dimensional EGB gravity on $(D-4)$-dimensional maximally symmetric space, and then make the replacement of the Gauss-Bonnet coupling $\alpha\rightarrow\frac{\alpha}{D-4}$ and take the $D\rightarrow4$ limit.
Through this procedure, an extra scalar degree of freedom is introduced in addition to the spin-2 degree of freedom, the resulting model is a special scalar-tensor theory that belongs to the family of Horndeski gravity. Other attempts involve the method of conformally rescaling the metric, then subtracting the original action from the new one associated with the rescaled metric, and taking the $D\rightarrow4$ limit at last\cite{2004.08362,2004.09472}. The action obtained via this method is compatible with the one obtained via the Kaluza-Klein method.

We intend to investigate black hole solution of the 4D Horndeski  gravity  coupled to Born-Infeld (BI) electromagnetic fields.  BI electrodynamics was proposed originally with the motivation of obtaining a finite value of the self-energy of electron\cite{BI}. In Ref.\cite{Fradkin:1985qd}, the authors showed that BI action arises naturally from string theory. The D3-brane dynamics was also noticed to be governed by BI action\cite{Tseytlin:1986ti}. In recent years, BI theory has been vastly used to study dark energy, holographic superconductor, holographic entanglement entropy, and holographic complexity \cite{0307177,1508.05926,1401.6505,1810.02208}, etc. Black hole solutions have been constructed for BI electromagnetic fields coupled to Einstein gravity\cite{0306120,0406169,0410158}, Gauss-Bonnet gravity\cite{0408078}, higher-order Lovelock gravity\cite{0802.2637,1712.08798,1804.10951}, and massive gravity\cite{1508.01311,1608.03148}.  Thermodynamics of the BI black holes have been studied in\cite{0410158,0408078,0802.2637,1712.08798,1804.10951,1508.01311,1608.03148,1011.3184,1311.7299,1401.0785}. In this paper, we construct novel dyonic black hole solution of 4D Horndeski gravity coupled to BI electromagnetic fields, and study thermodynamics of the black hole.

The paper is organized as, in section \ref{section2},  the action of 4D Horndeski gravity coupled to BI electromagnetic fields is presented, and the dyonic BI black hole solution is given. In section \ref{section3}, the first law of thermodynamics is checked, and the thermal phase transitions of the black holes with different spatially topologies are studied in extended phase space. We summarize our results in the last section.

\section{Black hole solution\label{section2}}
\subsection{Horndeski gravity in four dimensions}
In general $D$ dimensions, the action of EGB gravity coupled to BI electromagnetic fields is given by
\begin{align}\label{oldaction}
I_D=\frac{1}{16\pi G_D}\int d^Dx\sqrt{-g}\left(R-2\Lambda+\alpha\mathcal{L}_{GB}+16\pi G_D L(F)\right),
\end{align}
where $\mathcal{L}_{GB}$ is the Gauss-Bonnet density
\begin{equation}
\mathcal{L}_{GB}=R_{\mu\nu\rho\sigma}R^{\mu\nu\rho\sigma}-4R_{\mu\nu}R^{\mu\nu}+R^2,
\end{equation}
and $L(F)$ takes the form
\begin{align}\label{BI}
L(F)=\beta^2\sqrt{-\mathrm{det}(g_{\mu\nu})}-\beta^2\sqrt{-\mathrm{det}\left(g_{\mu\nu}+\frac{F_{\mu\nu}}{\beta}\right)},
\end{align}
$F_{\mu\nu}=\partial_\mu A_\nu-\partial_\nu A_\mu$ is the field strength tensor. Note that (\ref{BI}) tends to the Maxwell Lagrangian $-\frac 14 F_{\mu\nu}F^{\mu\nu}$ in the limit $\beta\rightarrow\infty$.

To obtain the 4D theory of gravity, one considers the Kaluza-Klein diagonal reduction of the action (\ref{oldaction}), with metric ansatz
\begin{equation}
ds_D^2=ds_p^2+e^{2\phi}d\Sigma_{D-p,\lambda}^2,
\end{equation}
where the breathing scalar $\phi$ depends only on the external $p$-dimensional coordinates. The line elements $d\Sigma_{D-p,\lambda}^2$ describe the internal maximally symmetric space, and $\lambda$ denotes the sign of the Euclidean space curvature. After the Kaluza-Klein procedure action (\ref{oldaction}) reduces to the $p$-dimensional action\cite{2003.11552,2004.09472}
\begin{align}\label{actionp1}
I_p=\frac{1}{16\pi G_p}\int & d^px\sqrt{-g}e^{(D-p)\phi}\Big\{R
-2\Lambda_0+16\pi G_p L(F)+(D-p)(D-p-1)\big((\partial\phi)^2+\lambda e^{-2\phi}\big)\nonumber\\
&
+\alpha\Big(\mathcal{L}_{GB}-2(D-p)(D-p-1)\left[2G^{\mu\nu}\partial_\mu\phi\partial_\nu\phi-\lambda Re^{-2\phi}\right]\nonumber\\
& -(D-p)(D-p-1)(D-p-2)\left[2(\partial\phi)^2\Box\phi+(D-p-1)((\partial\phi)^2)^2\right]\nonumber\\
&+(D-p)(D-p-1)(D-p-2)(D-p-3)\left[2\lambda (\partial\phi)^2e^{-2\phi}+\lambda^2e^{-4\phi}\right]\Big)\Big\}\,,
\end{align}
where $G_{\mu\nu}$ is Einstein tensor. For $p\leq4$, it is free to add
\begin{align}\label{actionadd}
-\frac{\alpha}{16\pi G_p}\int d^px\sqrt{-g}\, \mathcal{L}_{GB}
\end{align}
to action (\ref{actionp1}) without affecting the field equations, since (\ref{actionadd}) is just a topological invariant.
Now rescaling the Gauss-Bonnet coupling as $\alpha\rightarrow\frac{\alpha}{D-p}$ and taking the $D\rightarrow p$ limit, one obtains the $p$-dimensional theory
\begin{align}\label{action}
I_{p}&=\frac{1}{16\pi G_p}\int d^px\sqrt{-g}\left[R-2\Lambda+16\pi G_pL(F)+\alpha\left(-2\lambda R e^{-2\phi}
-12\lambda(\partial\phi)^2e^{-2\phi}\right.\right.\nonumber\\
&\left.\left.-6\lambda^2e^{-4\phi}+\phi\mathcal{L}_{GB}+4G^{\mu\nu}\partial_{\mu}\phi\partial_{\nu}\phi-4(\partial\phi)^2\square\phi+2((\partial\phi)^2)^2\right)\right].
\end{align}
This is the Horndeski gravity coupled to BI electromagnetic fields. This theory is well defined, and the entropy obtained by Wald formula is free from singularity.

\subsection{equations of motion}
Variation with respect to the electromagnetic field gives rise to the field equation
\begin{align}\label{eomEM}
\mathcal{E}_{A}=\nabla_\mu\left[\frac{\sqrt{-h}}{\sqrt{-g}}\beta(h^{-1})^{[\mu\nu]}\right]=0
\end{align}
where $h_{\mu\nu}\equiv g_{\mu\nu}+\frac{F_{\mu\nu}}{\beta}$, and $h\equiv \mathrm{det} (h_{\mu\nu})$. The symmetric part and antisymmetric part of $h_{\mu\nu}$ are denoted respectively by  $h_{(\mu\nu)}$ and $h_{[\mu\nu]}$. $(h^{-1})^{\mu\nu}$ denotes the inverse of $h_{\mu\nu}$, similarly, $(h^{-1})^{(\mu\nu)}$ and $(h^{-1})^{[\mu\nu]}$ are  the symmetric and antisymmetric parts of $(h^{-1})^{\mu\nu}$ respectively. \\
The equation of motion of $\phi$ is given by\cite{2004.09472,2011.08604}
\begin{align}
\mathcal{E}_{\phi}=&-\mathcal{L}_{GB}+8G^{\mu\nu}\nabla_{\nu}\nabla_{\mu}\phi+8R^{\mu\nu}\nabla_{\mu}\phi\nabla_{\nu}\phi-8(\square\phi)^2+8(\nabla\phi)^2\square\phi\nonumber\\
&+16\nabla^{\mu}\phi\nabla^{\nu}\nabla_{\nu}\nabla_{\mu}\phi+8\nabla_{\mu}\nabla_{\nu}\nabla^{\mu}\nabla^{\nu}\phi-24\lambda^2e^{-4\phi}-4\lambda Re^{-2\phi}\nonumber\\
&+24\lambda e^{-2\phi}\left((\nabla\phi)^2-\square\phi\right)=0,
\end{align}
The variation with respect to the metric yields
\begin{align}
{\mathcal E}_{\mu\nu} =& \Lambda g_{\mu\nu} +  G_{\mu\nu}-8\pi G\beta^2 g_{\mu\nu}+8\pi G\beta^2\frac{\sqrt{-h}}{\sqrt{-g}} h_{(\mu\nu)}\nonumber\\
&+ \alpha \bigg[\phi H_{\mu\nu}  -2 R \left[(\nabla_{\mu} \phi)(\nabla_{\nu} \phi) + \nabla_{\nu} \nabla_{\mu} \phi \right] + 8 R_{({\mu}}^{\rho} \nabla_{{\nu})} \nabla_{\rho} \phi + 8 R_{({\mu}}^{\rho} (\nabla_{{\nu})}\phi) (\nabla_{\rho} \phi)
\nonumber\\
&- 2 G_{\mu\nu} \left[(\nabla \phi)^2 +  2\Box \phi \right]   - 4 \left[ (\nabla_{\mu} \phi)(\nabla_{\nu} \phi) + \nabla_{\nu} \nabla_{\mu} \phi \right] \Box \phi + 3 \lambda^2 e^{-4 \phi} g_{\mu\nu}
\nonumber\\
&+ 8 (\nabla_{({\mu}} \phi) (\nabla_{{\nu})} \nabla_{\rho} \phi ) \nabla^{\rho} \phi - 4 g_{\mu\nu} R^{\rho\sigma} \left[\nabla_{\rho} \nabla_{\sigma} \phi + (\nabla_{\rho} \phi)(\nabla_{\sigma} \phi) \right] + 2 g_{\mu\nu} (\Box \phi)^2 \nonumber\\
&- 2 g_{\mu\nu} (\nabla_{\rho} \nabla_{\sigma} \phi)(\nabla^{\rho} \nabla^{\sigma} \phi)
- 4 g_{\mu\nu} (\nabla^{\rho} \phi ) (\nabla^{\sigma} \phi) (\nabla_{\rho} \nabla_{\sigma} \phi) + 4 (\nabla_{\rho} \nabla_{\nu} \phi)(\nabla^{\rho} \nabla_{\mu} \phi) \nonumber\\
&+ 4 R_{\mu\rho\nu\sigma} \left[(\nabla^{\rho} \phi)(\nabla^{\sigma} \phi) + \nabla^{\sigma} \nabla^{\rho} \phi \right]
-\left[g_{\mu\nu}(\nabla \phi)^2 -4(\nabla_{\mu} \phi)(\nabla_{\nu} \phi) \right](\nabla \phi)^2\nonumber\\
&- 2 \lambda e^{-2 \phi} \left( G_{\mu\nu} + 2 (\nabla_{\mu} \phi)(\nabla_{\nu} \phi) + 2 \nabla_{\nu} \nabla_{\mu} \phi - 2 g_{\mu\nu} \Box \phi + g_{\mu\nu} (\nabla \phi)^2 \right)  \bigg]=0.
\end{align}
Here and in the following we label the gravitational constant $G_p$ as $G$ for simplicity.

Combining the last two equations in the following manner yields
\begin{align}\label{eom}
g^{\mu\nu}\mathcal{E}_{\mu\nu}+\frac{\alpha}{2}\mathcal{E}_{\phi}=
4\Lambda-R-\frac{\alpha}2\mathcal{L}_{GB}-32\pi G\beta^2+8\pi G\beta^2\frac{\sqrt{-h}}{\sqrt{-g}} h_{(\mu\nu)}g^{\mu\nu}=0,
\end{align}
which is independent of the breathing scalar $\phi$ and curvature of the internal space.

\subsection{black hole solution}
To solve the field equations, we assume $\phi=\phi(r)$, and take the metric and field strength ansatz as
\begin{align}
ds_4^2&=-e^{-2\chi(r)}f(r)dt^2+\frac 1{f(r)}dr^2+r^2\left(\frac{du^2}{1-ku^2}+(1-ku^2)d\varphi^2\right),\label{metric}\\
F&=-a'(r)dt\wedge dr+p du\wedge d\varphi.\label{strength2}
\end{align}
Substituting (\ref{metric}) and (\ref{strength2}) into (\ref{eomEM}), and making $\chi(r)$ to be zero, one has
\begin{align}
a'(r)=\frac{q\beta}{\sqrt{p^2+q^2+\beta^2 r^4}}.
\end{align}
Note that, unlike Maxwell theory, the infinity in the intensity at $r=0$ has been removed, thereby the infinity in the potential at $r=0$ is absent too.
Now combining (\ref{eom}) together with (\ref{metric}), one obtains
\begin{align}\label{eom1}
&-\frac{2 \alpha  \left(f'^2+(f-k) f''\right)}{r^2}+\frac{4 f'}{r}+f''+\frac{-2k+2 f}{r^2}+4 \Lambda\nonumber\\
&-16 \pi G \beta r^{-2}\left(2\beta r^2-(p^2+q^2+2\beta^2 r^4)\left(p^2+q^2+\beta^2 r^4\right)^{-1/2}\right)=0,
\end{align}
where we denote $\frac{df(r)}{dr}$, $\frac{d^2f(r)}{dr^2}$ as $f'$, $f''$ for short. This equation is not enough  to find the explicit form of $f(r)$, one has to find other field equations.  Substituting the metric ansatz (\ref{metric}) into (\ref{action}), and discarding the total derivative terms, one obtains the effective Lagrangian\cite{2003.11552}
\begin{eqnarray}\label{lag}
\mathcal{L}&=&\frac{e^{-\chi}}{6r^2}\left[-3 \left(4 r^3 f'+4 \Lambda  r^4+4r^2f-4kr^2+8\alpha k r^2f\phi'^2\right)\right.\nonumber\\
&&+24\alpha k r^2\phi'(f'-2f\chi ')-8 \alpha r^2 f    f' \phi ' \left(r^2 \phi '^2-3 r \phi '+3\right)\nonumber\\
&&+4 \alpha  r^2 f^2 \phi ' \left(4 \chi ' \left(r^2 \phi '^2-3 r \phi '+3\right)+\phi ' \left(3 r^2 \phi '^2-8 r \phi '+6\right)\right)\nonumber\\
&&+24\alpha \lambda r^2e^{-2\phi}\left(r^2f'\phi'-2r^2f\chi'\phi'-3r^2f\phi'^2+r f'+f-k\right)\nonumber\\
&&\left.-36\alpha \lambda^2 r^4 e^{-4\phi}+96\pi G\left(\beta^2r^4-\beta r^2(p^2+\beta^2r^4)(p^2+q^2+\beta^2r^4)^{-1/2}\right)\right].
\end{eqnarray}
Taking variation of (\ref{lag}) with respect to $f(r)$, and making $\chi(r)$ to be zero, one has
\begin{equation}
\left(-r^2\lambda+e^{2\phi}(-k+f(r\phi'-1)^2)\right)\left(\phi '^2+\phi ''\right)=0,
\end{equation}
which implies
\begin{equation}\label{varif}
f(r)= \frac{k+r^2\lambda e^{-2\phi}}{(r\phi'-1)^2}.
\end{equation}
The solution of $\phi$ is given by
\begin{align}
\phi=\log(r)+\log\left(\cosh(\sqrt{k}\psi)\pm\sqrt{1+\lambda/k}\sinh(\sqrt{k}\psi)\right),\psi=\int_{r_{+}}^r\frac{du}{u\sqrt{f(u)}}.
\end{align}
Variation of the effective action with respect to $\chi$ yields an equation that is rather complicated
\begin{align}\label{varochi}
&2e^{-4\phi-\chi}(2\alpha\lambda e^{2\phi}(-k+r f'(1-r\phi')+f(1-4r\phi'+r^2(\phi'^2-2\phi'')))\nonumber\\
&+e^{4\phi}(k-r(f'+r\Lambda)-2k\alpha
f'\phi'+f(-1+2\alpha f'\phi'(3-3r\phi'+r^2\phi'^2)\nonumber\\
&-2k\alpha(\phi'^2+2\phi''))+\alpha f(r^2\phi'^4+4\phi'^2-8r\phi'\phi''+\phi'^2(-2+4r^2\phi'^2))))\nonumber\\
&-16\pi Ge^{-\chi}\left(\beta^2r^2-\beta(p^2+\beta^2r^4)(p^2+q^2+\beta^2r^4)^{-1/2}\right).
\end{align}
Fortunately, with (\ref{varif}) the equation (\ref{varochi}) can be simplified greatly to be
\begin{align}\label{eom2}
&r \left(r^2-2 \alpha  f(r)+2k\alpha\right) f'(r)+f(r) \left(-2\alpha k+\alpha  f(r)+r^2\right)+\Lambda  r^4\nonumber\\
&+\alpha k^2-kr^2+8\pi G\left(-\beta^2r^4+\beta r^2\sqrt{p^2+q^2+\beta^2r^4}\right)=0
\end{align}
Combining (\ref{eom1}) and (\ref{eom2}) one is now able to give the exact black hole solution:
\begin{align}\label{solution}
f(r)=&k+\frac{r^2}{2\alpha}\left(1-\left(1+\frac{32\pi G\alpha M}{\Sigma_2 r^3}-\frac{32\pi G\alpha\beta^2}{3}+\frac{4\alpha\Lambda}{3}+\frac{32\pi G\alpha\beta^2}{3}\right.\right.\nonumber\\
&\left.\left.\cdot\left(\sqrt{1+\frac{p^2+q^2}{\beta^2r^4}}-\frac{2(p^2+q^2)}{\beta^2r^4}\sideset{_2}{_1}{\mathop{F}}\left[\frac{1}{4},\frac{1}{2},
\frac{5}{4},-\frac{p^2+q^2}{\beta^2r^4}\right]\right)\right)^{1/2}\right),
\end{align}
where $\Sigma_2$ is the spatial 2-volume.

In order to study the behavior of $f(r)$, we expand $f(r)$ in the small-$r$ and large-$r$ regions respectively, yielding
\begin{align}\label{f0}
f(r)=&k-\sqrt{8\pi G M/(\alpha\Sigma_2)-16G\sqrt{\beta\pi}(p^2+q^2)^{3/4}\Gamma(1/4)\Gamma(5/4)/(3\alpha)}r^{1/2}\nonumber\\
&-\frac{\frac{8}{3}\pi G\beta\sqrt{p^2+q^2}+\frac{64}{3}\pi G\beta\sqrt{p^2+q^2}\Gamma(5/4)/\Gamma(1/4)}{\sqrt{32\pi G \alpha M/\Sigma_2-64\alpha G\sqrt{\beta\pi}(p^2+q^2)^{3/4}\Gamma(1/4)\Gamma(5/4)/3}}r^{3/2}+\frac{r^2}{2\alpha}+\mathcal{O}(r)^{5/2},\\
f(r)&=k+\frac{3-\sqrt{9+12\alpha\Lambda}}{6\alpha}r^2-\frac{8\pi GM}{\Sigma_2\sqrt{1+4\alpha\Lambda/3}}r^{-1}+\mathcal{O}(r)^{-2}.\label{finfty}
\end{align}
From (\ref{f0}) one learns that,  $f(r)$ is finite when $r\rightarrow0$. This property is specific for BI black holes, i.e., for BI black holes the metric may be free from divergence at the origin  while the curvature invariants definitely diverge there. $f(r)$ is finite at the origin originates partly from the nonlinearity of matter fields, partly from the model of gravity theory, and partly from the dimensions of  spacetime. We only consider the case that the dimensions of the spacetime is  not less than 4. For the black holes in Einstein-Born-Infeld gravity, $f(r)$ diverges at the origin.   For the black holes in Gauss-Bonnet-Born-Infeld gravity,  $f(r)$ is finite in 5 dimensions  while it diverges in higher than 5 dimensions at the origin. For the black holes in Gauss-Bonnet-Maxwell gravity, $f(r)$ diverges at $r=0$. For the black holes in 3rd order Lovelock gravity coupled to BI electromagnetic fields, $f(r)$ is finite in 7 dimensions  while it diverges in higher dimensions at $r=0$.

One also learns from (\ref{f0}) that, in order to ensure $f(r)$ to be well defined, the black hole mass $M$ is necessary to be larger than some critical one
\begin{align}
M_c=\frac{2\sqrt{\beta}}{3\sqrt{\pi}}(p^2+q^2)^{3/4}\Gamma(1/4)\Gamma(5/4)\Sigma_2,
\end{align}
which is independent of $k$. In order to preserve $f(r)>0$ in the large-$r$ region, from (\ref{finfty}) we know $\Lambda$ has to be negative, i.e., we only consider the AdS black holes.

\begin{figure}[h]
\begin{center}
\includegraphics[width=.32\textwidth]{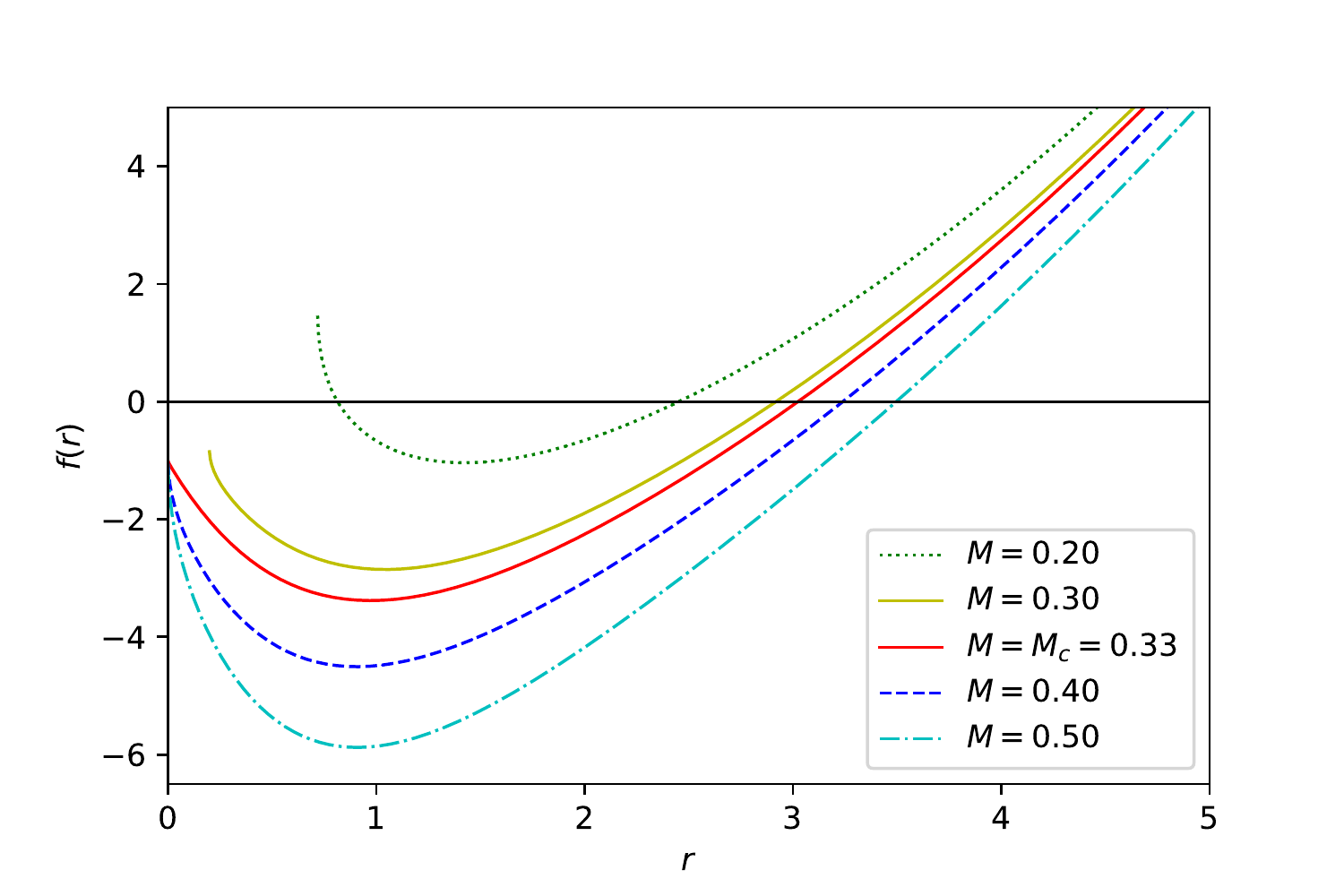}
\includegraphics[width=.32\textwidth]{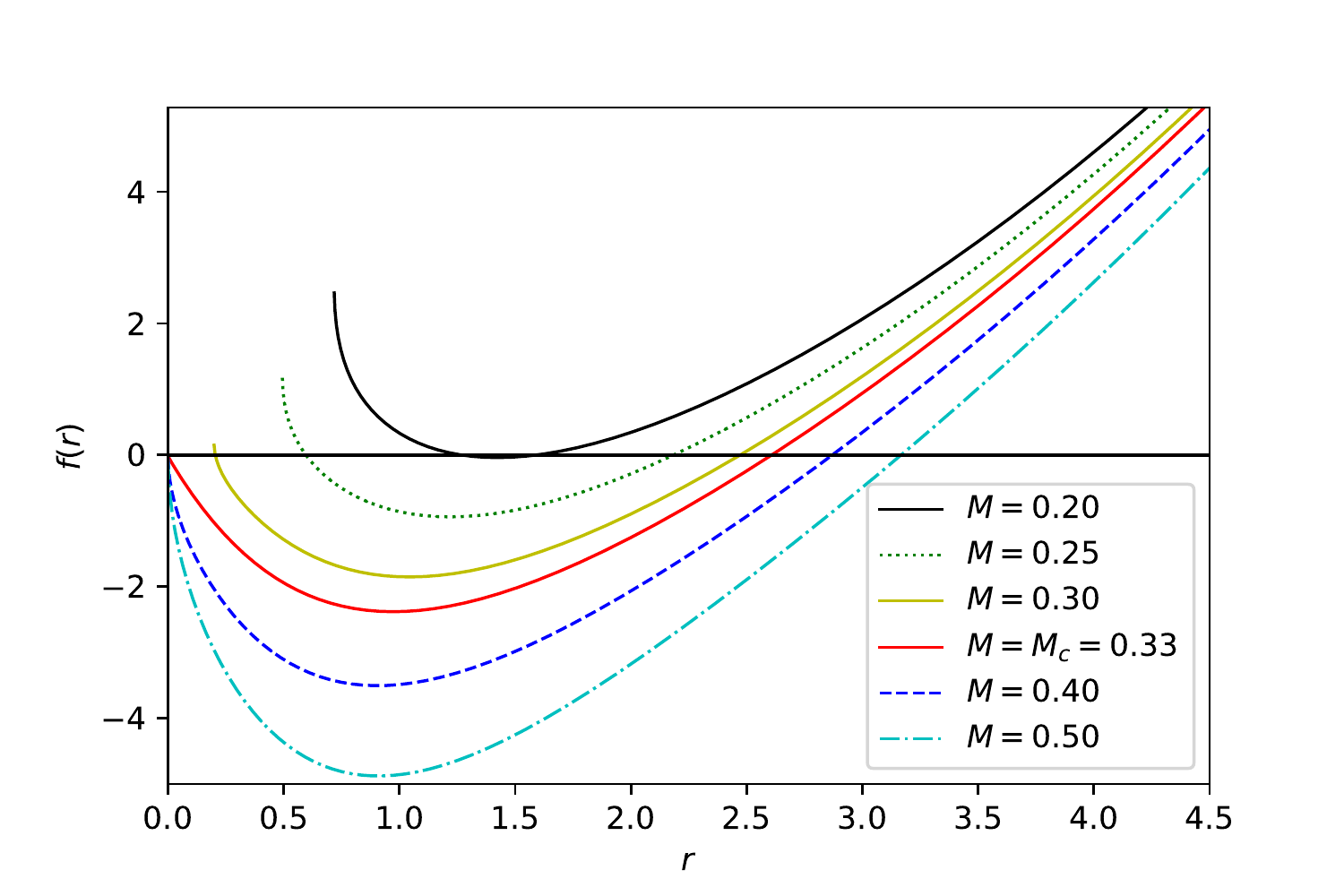}
\includegraphics[width=.32\textwidth]{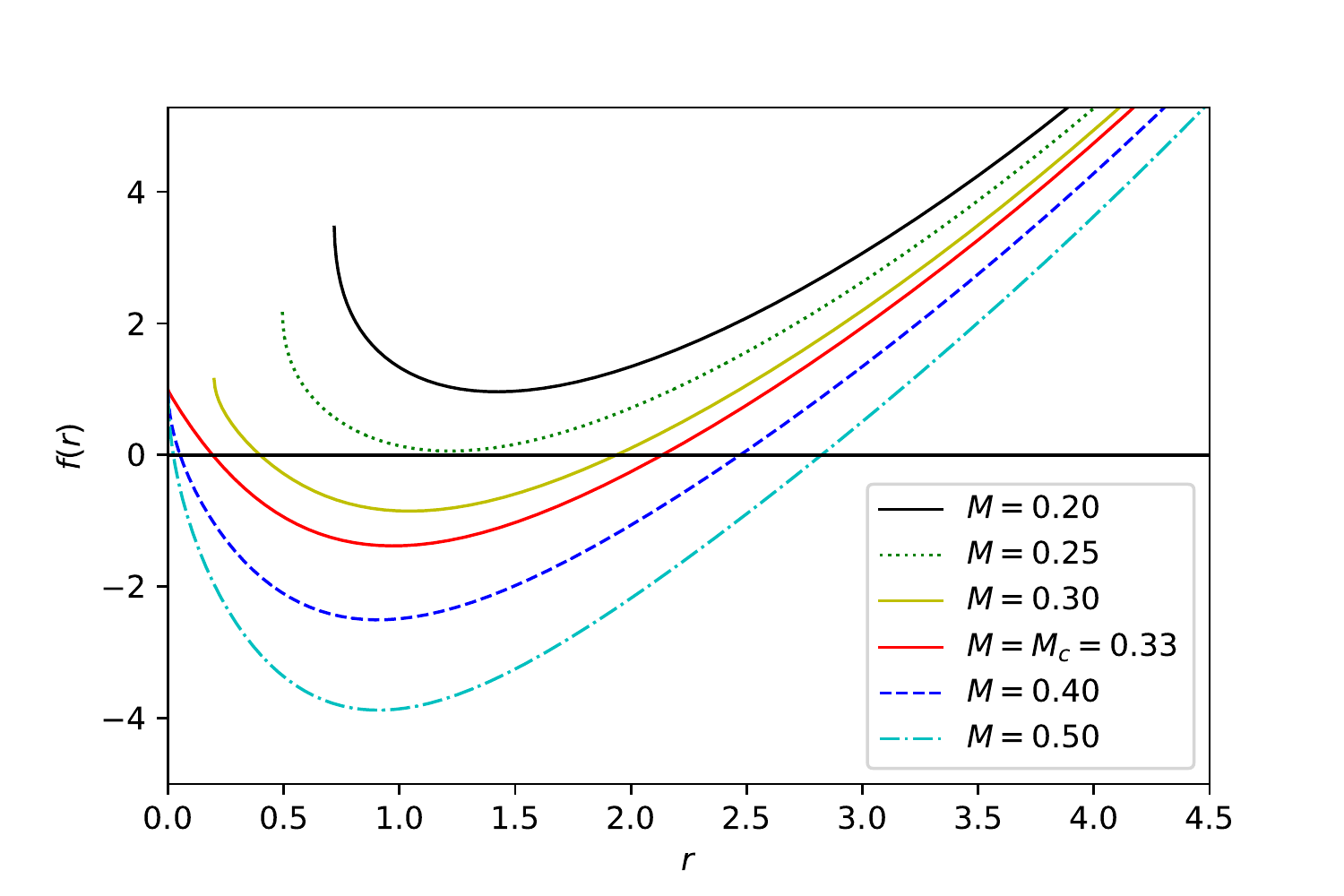}
\end{center}
\caption{The metric function $f(r)$ for the parameters selection $G=1, \beta=0.2, p=0.5, q=0.5, \Lambda=-1, \alpha=0.1$. From left to right, the three plots correspond to $k=-1,0,+1$ respectively. The red line on each plot denotes the case $M=M_c$.}
\label{fig1}
\end{figure}

In Fig.\ref{fig1} we present the behaviors of $f(r)$. One sees from Fig.\ref{fig1} that, for $M>M_c$, the planar and hyperbolic black holes possess single horizon.  The inner (Cauchy) horizon of the BI black hole turns into the curvature singularity due to perturbatively instability\cite{1702.06766}.  While in Einstein gravity, the planar and hyperbolic BI black holes may possess more than one horizon for some parameter selections. As shown in Fig.\ref{fig1}, the spherical black hole possesses double horizons.
For $M<M_c$, the black holes are not well defined in the whole spacetime\cite{0408078}.


The behavior of $f(r)$ implies that, there exists no extremal black hole for our case. We take the $k=0$ case as an example to illustrate this analytically. We substitute
\begin{align}
r_e\equiv\sqrt{8\pi G\beta}\left(\frac{p^2+q^2}{\Lambda^2-16\pi G\Lambda\beta^2}\right)^{1/4},
\end{align}
which is obtained by setting the temperature (\ref{temperature}) to be zero, into the black hole mass (\ref{mass}), and replace the hypergeometric function with 1 since $0<\sideset{_2}{_1}{\mathop{F}}\left[\frac{1}{4},\frac{1}{2},
\frac{5}{4},-\frac{p^2+q^2}{\beta^2r^4}\right]<1$, then the mass is given by
\begin{align}
M_e\equiv-\frac{(p^2+q^2)^{3/4}\Sigma_2\left(64\pi^2G^2\beta^4-24\pi G\Lambda\beta^2+\Lambda^2\right)}{3\sqrt{2\pi G\beta\Lambda(-16\pi G\beta^2+\Lambda)}},
\end{align}
which is definitely negative for $\Lambda<0$. Therefore, extremal AdS black holes do not exist for the case we discussed.

\section{Thermodynamics\label{section3}}
In this section, we check first law of thermodynamics and study thermodynamic phase transitions of the black holes in extended phase space. First Let's give the thermodynamic quantities.
Mass of the black hole is given by
\begin{align}\label{mass}
M=&\frac{\Sigma_2}{24\pi Gr_{+}}\left(3kr_{+}^2+3k^2\alpha+8\pi G\beta r_{+}^2\left(\beta r_{+}^2-\sqrt{p^2+q^2+\beta^2 r_{+}^4}\right)\nonumber\right.\\
&\left.-\Lambda r_{+}^4+16\pi G(p^2+q^2)
\sideset{_2}{_1}{\mathop{F}}\left[\frac{1}{4},\frac{1}{2},
\frac{5}{4},-\frac{p^2+q^2}{\beta^2r_{+}^4}\right]\right),
\end{align}
where $r_{+}$ is the outermost horizon of the black hole. Temperature of the black hole reads
\begin{align}\label{temperature}
T=\frac{-k^2\alpha^2+r_{+}^4(-1-\alpha\Lambda+8\pi G\alpha\beta^2)+r_{+}^2\left(r_{+}^2+k\alpha-8\pi G\alpha\beta\sqrt{p^2+q^2+r_{+}^4\beta^2}\right)}{4\alpha \pi r_{+}(r_{+}^2+2k\alpha)}
\end{align}
With the Iyer-Wald formula, entropy of the black hole is given by
\begin{align}\label{entropy1}
S&=-2\pi\oint d^{2}x\sqrt{\gamma}Y^{\mu\nu\rho\sigma}\epsilon_{\mu\nu}\epsilon_{\rho\sigma}\nonumber\\
&=-\frac{1}{8G}\oint d^{2}x\sqrt{\gamma}\left(-2-4\alpha\phi\tilde{R}(\gamma)+\alpha\delta^\rho_{[\mu}\partial_{\nu]}\phi\partial^\sigma\phi\epsilon^{\mu\nu}\epsilon_{\rho\sigma}
-\frac{1}{2}\alpha(\partial\phi)^2\delta^\rho_{[\mu}\delta^\sigma_{\nu]}\epsilon^{\mu\nu}\epsilon_{\rho\sigma}\right),
\end{align}
where the first term in the bracket comes from Einstein gravity, the second term comes from the term $\alpha\phi\mathcal{L}_{GB}$ in  action (\ref{action}), while the third and fourth terms come from $4\alpha G^{\mu\nu}\partial_\mu\phi\partial_\nu\phi$ in the action.  Straightforward calculations show that the last two terms in (\ref{entropy1}) cancel out, i.e., the $4\alpha G^{\mu\nu}\partial_\mu\phi\partial_\nu\phi$ term in the action does not contribute to entropy of the black hole. Only the first two terms contribute to the entropy, which gives
\begin{align}\label{entropy}
S=\frac{\Sigma_2(r_{+}^2+4k\alpha\log(r_{+}))}{4G}.
\end{align}

The electric and magnetic charges are given by
\begin{align}
Q_e=\Sigma_2\sqrt{-h}\left(h^{-1}\right)^{[tr]}\big|_{r\rightarrow\infty}=q\Sigma_2,  \;\;\;\;\;\;\;Q_m=\Sigma_2F_{xy}|_{r\rightarrow\infty}=p\Sigma_2.
\end{align}
Note the above electric charge as a conserved quantity follows from the equation of
motion (\ref{eomEM}). The electric and magnetic potentials are given by
\begin{align}
\Phi_e=\frac{q}{r_{+}}\sideset{_2}{_1}{\mathop{F}}\left[\frac{1}{4},\frac{1}{2},
\frac{5}{4},-\frac{p^2+q^2}{\beta^2r_{+}^4}\right],  \;\;\;\;\;\;\;\Phi_m=\frac{p}{r_{+}}\sideset{_2}{_1}{\mathop{F}}\left[\frac{1}{4},\frac{1}{2},
\frac{5}{4},-\frac{p^2+q^2}{\beta^2r_{+}^4}\right],
\end{align}
In extended phase space, the  thermodynamic pressure of the system is identified as\cite{1205.0559,1208.6251}
\begin{align}\label{pressure}
P=-\frac{\Lambda}{8\pi G}.
\end{align}
The thermodynamic volume conjugate to $P$ is given by
\begin{align}
V=\frac{r_{+}^3\Sigma_2}{3}
\end{align}
With all the thermodynamic quantities given above, it's straightforward to check the first law of thermodynamics
\begin{align}
\delta M=T\delta S+V\delta P+\Phi_e\delta Q_e+\Phi_m\delta Q_m
\end{align}
is satisfied.

\begin{figure}[h]
\begin{center}
\includegraphics[width=.60\textwidth]{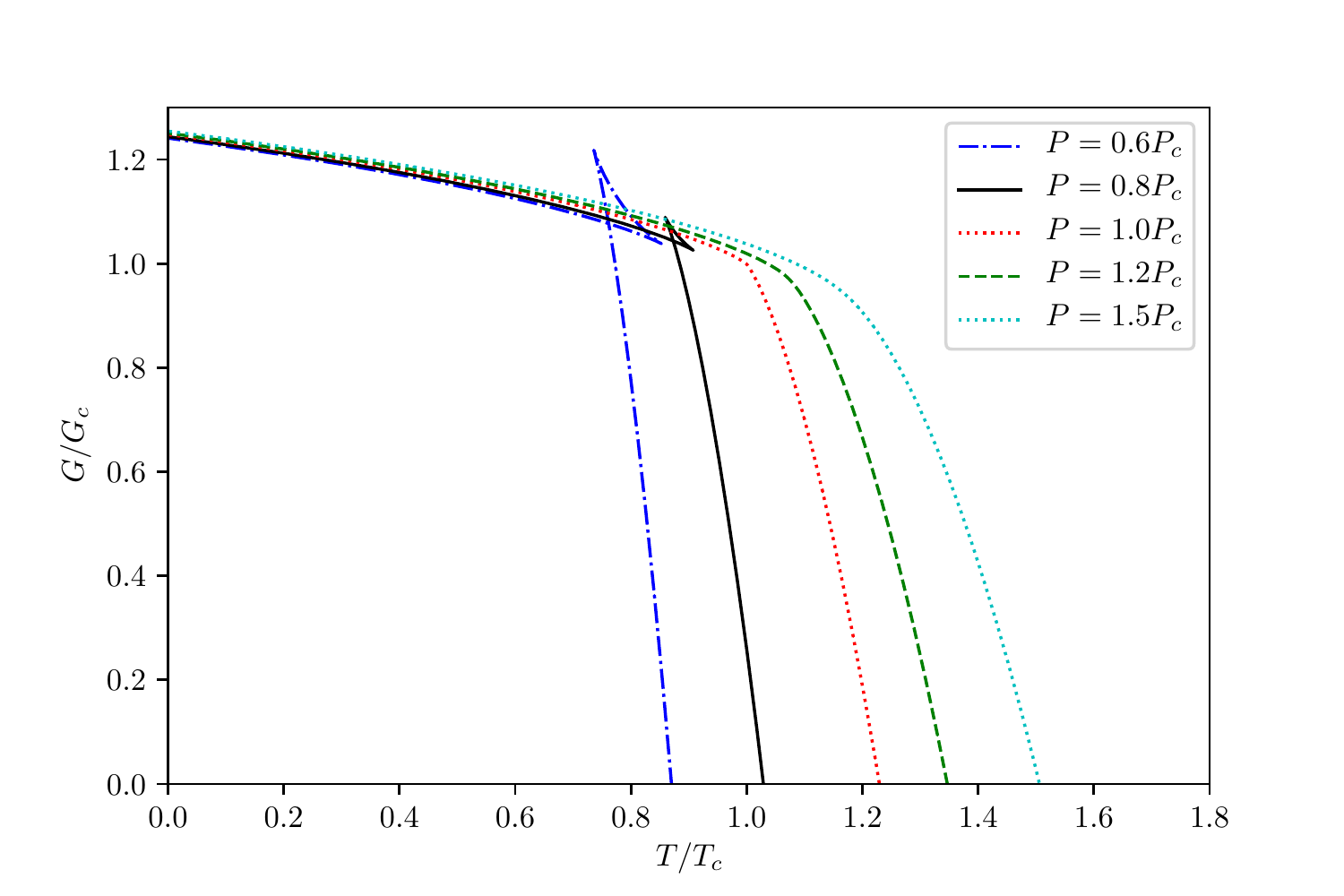}
\end{center}
\caption{$G$-$T$ plot for the parameters  fixed as $G=1, \alpha=0.1, \beta=10, p=0.2, q=0.2$.}
\label{fig2}
\end{figure}

Now let's examine if there exist thermal phase transitions of the black hole. The critical point is determined by the equations
\begin{align}
\frac{\partial P}{\partial r_+}\bigg|_{r_+=r_c,T=T_c}=\frac{\partial^2 P}{\partial r_+^2}\bigg|_{r_+=r_c,T=T_c}=0.
\end{align}
From (\ref{temperature}) one can solve out $\Lambda$ in terms of $T$, and substitute $\Lambda$ into the definition of pressure (\ref{pressure}), yielding
\begin{align}
P=\frac{k^2\alpha-kr_{+}^2+T(4\pi r_{+}^3+8k\pi\alpha r_{+})-8\pi G\beta^2r_{+}^4+8\pi G\beta r_{+}^2\sqrt{p^2+q^2+\beta^2r_{+}^4}}{8\pi Gr_{+}^4}.
\end{align}
From the equation $\frac{\partial P}{\partial r_+}=0$, $T$ can be solved out. Substituting this $T$ into $\frac{\partial^2 P}{\partial r_+^2}$ one obtains the final expression of
$\frac{\partial^2 P}{\partial r_+^2}$, which is a little lengthy and will not be presented here.
While, for $k=0$ the expression is quite simple
\begin{align}
\frac{\partial^2 P}{\partial r_+^2}=\frac{2\beta(p^2+q^2)(p^2+q^2+3r^4\beta^2)}{r^4(p^2+q^2+r^4\beta^2)^{3/2}},
\end{align}
which can't be zero. Therefore, no phase transition exists for the planar black hole. For $k=-1$, the critical equations can be solved formally. However, if the solutions are resubstituted into the related quantities, one finds that either $f(r)$ becomes imaginary or the entropy becomes negative. Thus, in this case the black hole is thermodynamically stable and no phase transition exists either. For $k=+1$, the critical equation can be solved out numerically, in this case there exists thermodynamic phase transition  as shown in Fig.\ref{fig2}, from which one sees that the phase transition is the van der Waals-like one-order phase transition.

For different parameter selections, the results are listed in the table. From the table, one learns that either $\alpha$ or $\beta$ increases, both the critical temperature and the critical pressure decrease.\\

\begin{table}
\centering
\begin{tabular}{cccc}
  \hline
  \hline
  $\alpha$ & $\beta$ & $T_c$ & $P_c$  \\
  \hline
  0.1 & 6 & 0.03797860 & 0.00259284 \\
  \hline
  0.1 & 10 & 0.03797840 & 0.00259280  \\
  \hline
  0.1 & 14 & 0.03797835 & 0.00259279  \\
  \hline
  0.08 & 10 & 0.03887130 & 0.00270822  \\
  \hline
  0.12 & 10 & 0.03714610 & 0.00248712  \\
  \hline
\end{tabular}
\caption{Critical temperatures and pressures for the parameters $p$ and $q$ fixed as $p=q=0.2$.}
\end{table}

\section{Conclusions}
In this paper, we construct novel dyonic BI black hole solution of a 4D Horndeski gravity which is obtained from higher-dimensional EGB gravity through the Kaluza-Klein process. The metric function $f(r)$is devoid of divergence at the origin, while the essential singularity still exists there. This is different from the 4D Einstein BI black hole, which is divergent at the origin. This property is determined by the nonlinearity of BI electrodynamics, the scalar-tensor theory and the dimensions of spacetime together. There exist some critical mass $M_c$, the black hole mass must be larger than the critical mass in order to be well defined in the whole spacetime. For the planar and hyperbolic black holes, there exist only one horizon. For the spherical black hole, there exists double horizons.

The thermodynamic quantities of the black hole are calculated, the first law is checked to be satisfied. 
The thermal phase transitions are studied in extended phase space. Through solving the $P$-$V$ critical equations, it's found that the planar and hyperbolic black holes are  thermodynamically stable, no thermodynamic phase transition occurs. While, for spherical black holes, the van der Waals-like one-order phase transition occurs.
\section*{Acknowledgment}
The work of KM and LZC is supported by the National Natural Science Foundation of China (No. 62005199), the Key Research and Development Plan of Shandong Province (No. 2019GGX101073), and the Natural Science Foundation of Shandong Province (Nos. ZR2020LLZ001 and ZR2019LLZ006). The work of JQZ is supported by Natural Science Foundation of Shandong Province (No. ZR2020KF017).

\providecommand{\href}[2]{#2}\begingroup
\footnotesize\itemsep=0pt
\providecommand{\eprint}[2][]{\href{http://arxiv.org/abs/#2}{arXiv:#2}}


\begin{thebibliography}{}

\bibitem{Lovelock:1971yv}
D.~Lovelock,
``The Einstein tensor and its generalizations,''
J. Math. Phys. \textbf{12}, 498-501 (1971)
doi:10.1063/1.1665613

\bibitem{Glavan:2019inb}
D.~Glavan and C.~Lin,
``Einstein-Gauss-Bonnet Gravity in Four-Dimensional Spacetime,''
Phys. Rev. Lett. \textbf{124}, no.8, 081301 (2020)
doi:10.1103/PhysRevLett.124.081301
[arXiv:1905.03601 [gr-qc]].


\bibitem{2004.14468}
K.~Yang, B.~M.~Gu, S.~W.~Wei and Y.~X.~Liu,
``Born\textendash{}Infeld black holes in 4D Einstein\textendash{}Gauss\textendash{}Bonnet gravity,''
Eur. Phys. J. C \textbf{80}, no.7, 662 (2020)
doi:10.1140/epjc/s10052-020-8246-6
[arXiv:2004.14468 [gr-qc]].

\bibitem{2003.07769}
S.~W.~Wei and Y.~X.~Liu,
``Testing the nature of Gauss-Bonnet gravity by four-dimensional rotating black hole shadow,''
[arXiv:2003.07769 [gr-qc]].

\bibitem{2006.07913}
Z.~C.~Lin, K.~Yang, S.~W.~Wei, Y.~Q.~Wang and Y.~X.~Liu,
``Equivalence of solutions between the four-dimensional novel and regularized EGB theories in a cylindrically symmetric spacetime,''
Eur. Phys. J. C \textbf{80}, no.11, 1033 (2020)
doi:10.1140/epjc/s10052-020-08612-5
[arXiv:2006.07913 [gr-qc]].

\bibitem{2003.14275}
S.~W.~Wei and Y.~X.~Liu,
``Extended thermodynamics and microstructures of four-dimensional charged Gauss-Bonnet black hole in AdS space,''
Phys. Rev. D \textbf{101}, no.10, 104018 (2020)
doi:10.1103/PhysRevD.101.104018
[arXiv:2003.14275 [gr-qc]].

\bibitem{2004.07934}
S.~J.~Yang, J.~J.~Wan, J.~Chen, J.~Yang and Y.~Q.~Wang,
``Weak cosmic censorship conjecture for the novel $4D$ charged Einstein-Gauss-Bonnet black hole with test scalar field and particle,''
Eur. Phys. J. C \textbf{80}, no.10, 937 (2020)
doi:10.1140/epjc/s10052-020-08511-9
[arXiv:2004.07934 [gr-qc]].

\bibitem{2003.02523}
M.~Guo and P.~C.~Li,
``Innermost stable circular orbit and shadow of the $4D$ Einstein\textendash{}Gauss\textendash{}Bonnet black hole,''
Eur. Phys. J. C \textbf{80}, no.6, 588 (2020)
doi:10.1140/epjc/s10052-020-8164-7
[arXiv:2003.02523 [gr-qc]].

\bibitem{2003.13068}
C.~Y.~Zhang, P.~C.~Li and M.~Guo,
``Greybody factor and power spectra of the Hawking radiation in the $4D$ Einstein\textendash{}Gauss\textendash{}Bonnet de-Sitter gravity,''
Eur. Phys. J. C \textbf{80}, no.9, 874 (2020)
doi:10.1140/epjc/s10052-020-08448-z
[arXiv:2003.13068 [hep-th]].

\bibitem{2004.03141}
C.~Y.~Zhang, S.~J.~Zhang, P.~C.~Li and M.~Guo,
``Superradiance and stability of the regularized 4D charged Einstein-Gauss-Bonnet black hole,''
JHEP \textbf{08}, 105 (2020)
doi:10.1007/JHEP08(2020)105
[arXiv:2004.03141 [gr-qc]].

\bibitem{2004.09214}
S.~Mahapatra,
``A note on the total action of 4D Gauss\textendash{}Bonnet theory,''
Eur. Phys. J. C \textbf{80}, no.10, 992 (2020)
doi:10.1140/epjc/s10052-020-08568-6
[arXiv:2004.09214 [gr-qc]].


\bibitem{2004.08362}
P.~G.~S.~Fernandes, P.~Carrilho, T.~Clifton and D.~J.~Mulryne,
``Derivation of Regularized Field Equations for the Einstein-Gauss-Bonnet Theory in Four Dimensions,''
Phys. Rev. D \textbf{102}, no.2, 024025 (2020)
doi:10.1103/PhysRevD.102.024025
[arXiv:2004.08362 [gr-qc]].

\bibitem{2009.13508}
M.~Gurses, T.~\c{C}.~\c{S}i\c{s}man and B.~Tekin,
``Comment on ''Einstein-Gauss-Bonnet Gravity in 4-Dimensional Space-Time'',''
Phys. Rev. Lett. \textbf{125}, no.14, 149001 (2020)
doi:10.1103/PhysRevLett.125.149001
[arXiv:2009.13508 [gr-qc]].

\bibitem{2004.12998}
J.~Arrechea, A.~Delhom and A.~Jim\'enez-Cano,
``Inconsistencies in four-dimensional Einstein-Gauss-Bonnet gravity,''
Chin. Phys. C \textbf{45}, no.1, 013107 (2021)
doi:10.1088/1674-1137/abc1d4
[arXiv:2004.12998 [gr-qc]].

\bibitem{2003.11552}
H.~Lu and Y.~Pang,
Phys. Lett. B \textbf{809}, 135717 (2020)
doi:10.1016/j.physletb.2020.135717
[arXiv:2003.11552 [gr-qc]].

\bibitem{2004.09472}
R.~A.~Hennigar, D.~Kubiz\v{n}\'ak, R.~B.~Mann and C.~Pollack,
``On taking the  $D\rightarrow4$  limit of Gauss-Bonnet gravity: theory and solutions,''
JHEP \textbf{07}, 027 (2020)
doi:10.1007/JHEP07(2020)027
[arXiv:2004.09472 [gr-qc]].

\bibitem{BI}
M.~Born and L.~Infeld,
``Foundations of the new field theory,''
Proc. Roy. Soc. Lond. A \textbf{144}, no.852, 425-451 (1934)
doi:10.1098/rspa.1934.0059

\bibitem{Fradkin:1985qd}
E.~S.~Fradkin and A.~A.~Tseytlin,
``Nonlinear Electrodynamics from Quantized Strings,''
Phys. Lett. B \textbf{163}, 123-130 (1985)
doi:10.1016/0370-2693(85)90205-9


\bibitem{Tseytlin:1986ti}
A.~A.~Tseytlin,
``Vector Field Effective Action in the Open Superstring Theory,''
Nucl. Phys. B \textbf{276}, 391 (1986)
[erratum: Nucl. Phys. B \textbf{291}, 876 (1987)]
doi:10.1016/0550-3213(86)90303-2

\bibitem{0307177}
E.~Elizalde, J.~E.~Lidsey, S.~Nojiri and S.~D.~Odintsov,
``Born-Infeld quantum condensate as dark energy in the universe,''
Phys. Lett. B \textbf{574}, 1-7 (2003)
doi:10.1016/j.physletb.2003.08.074
[arXiv:hep-th/0307177 [hep-th]].

\bibitem{1508.05926}
C.~Lai, Q.~Pan, J.~Jing and Y.~Wang,
``On analytical study of holographic superconductors with Born\textendash{}Infeld electrodynamics,''
Phys. Lett. B \textbf{749}, 437-442 (2015)
doi:10.1016/j.physletb.2015.08.014
[arXiv:1508.05926 [hep-th]].


\bibitem{1401.6505}
W.~Yao and J.~Jing,
``Holographic entanglement entropy in insulator/superconductor transition with Born-Infeld electrodynamics,''
JHEP \textbf{05}, 058 (2014)
doi:10.1007/JHEP05(2014)058
[arXiv:1401.6505 [hep-th]].

\bibitem{1810.02208}
K.~Meng,
``Holographic complexity of Born\textendash{}Infeld black holes,''
Eur. Phys. J. C \textbf{79}, no.12, 984 (2019)
doi:10.1140/epjc/s10052-019-7510-0
[arXiv:1810.02208 [hep-th]].

\bibitem{0306120}
S.~Fernando and D.~Krug,
``Charged black hole solutions in Einstein-Born-Infeld gravity with a cosmological constant,''
Gen. Rel. Grav. \textbf{35}, 129-137 (2003)
doi:10.1023/A:1021315214180
[arXiv:hep-th/0306120 [hep-th]].

\bibitem{0406169}
T.~K.~Dey,
``Born-Infeld black holes in the presence of a cosmological constant,''
Phys. Lett. B \textbf{595}, 484-490 (2004)
doi:10.1016/j.physletb.2004.06.047
[arXiv:hep-th/0406169 [hep-th]].

\bibitem{0410158}
R.~G.~Cai, D.~W.~Pang and A.~Wang,
``Born-Infeld black holes in (A)dS spaces,''
Phys. Rev. D \textbf{70}, 124034 (2004)
doi:10.1103/PhysRevD.70.124034
[arXiv:hep-th/0410158 [hep-th]].


\bibitem{0408078}
M.~Aiello, R.~Ferraro and G.~Giribet,
``Exact solutions of Lovelock-Born-Infeld black holes,''
Phys. Rev. D \textbf{70}, 104014 (2004)
doi:10.1103/PhysRevD.70.104014
[arXiv:gr-qc/0408078 [gr-qc]].

\bibitem{0802.2637}
M.~H.~Dehghani, N.~Alinejadi and S.~H.~Hendi,
``Topological Black Holes in Lovelock-Born-Infeld Gravity,''
Phys. Rev. D \textbf{77}, 104025 (2008)
doi:10.1103/PhysRevD.77.104025
[arXiv:0802.2637 [hep-th]].

\bibitem{1712.08798}
K.~Meng and D.~B.~Yang,
``Black holes of dimensionally continued gravity coupled to Born\textendash{}Infeld electromagnetic field,''
Phys. Lett. B \textbf{780}, 363-371 (2018)
doi:10.1016/j.physletb.2018.03.032
[arXiv:1712.08798 [gr-qc]].

\bibitem{1804.10951}
K.~Meng,
``Hairy black holes of Lovelock\textendash{}Born\textendash{}Infeld-scalar gravity,''
Phys. Lett. B \textbf{784}, 56-61 (2018)
doi:10.1016/j.physletb.2018.07.029
[arXiv:1804.10951 [gr-qc]].

\bibitem{1508.01311}
S.~H.~Hendi, B.~Eslam Panah and S.~Panahiyan,
``Einstein-Born-Infeld-Massive Gravity: adS-Black Hole Solutions and their Thermodynamical properties,''
JHEP \textbf{11}, 157 (2015)
doi:10.1007/JHEP11(2015)157
[arXiv:1508.01311 [hep-th]].

\bibitem{1608.03148}
S.~H.~Hendi, G.~Q.~Li, J.~X.~Mo, S.~Panahiyan and B.~Eslam Panah,
``New perspective for black hole thermodynamics in Gauss\textendash{}Bonnet\textendash{}Born\textendash{}Infeld massive gravity,''
Eur. Phys. J. C \textbf{76}, no.10, 571 (2016)
doi:10.1140/epjc/s10052-016-4410-4
[arXiv:1608.03148 [gr-qc]].

\bibitem{1011.3184}
D.~Zou, Z.~Yang, R.~Yue and P.~Li,
``Thermodynamics of Gauss-Bonnet-Born-Infeld black holes in AdS space,''
Mod. Phys. Lett. A \textbf{26}, 515-529 (2011)
doi:10.1142/S0217732311034724
[arXiv:1011.3184 [gr-qc]].

\bibitem{1311.7299}
D.~C.~Zou, S.~J.~Zhang and B.~Wang,
``Critical behavior of Born-Infeld AdS black holes in the extended phase space thermodynamics,''
Phys. Rev. D \textbf{89}, no.4, 044002 (2014)
doi:10.1103/PhysRevD.89.044002
[arXiv:1311.7299 [hep-th]].

\bibitem{1401.0785}
J.~X.~Mo and W.~B.~Liu,
``$P-V$ criticality of topological black holes in Lovelock-Born-Infeld gravity,''
Eur. Phys. J. C \textbf{74}, no.4, 2836 (2014)
doi:10.1140/epjc/s10052-014-2836-0
[arXiv:1401.0785 [gr-qc]].

\bibitem{2011.08604}
Y.~L.~Wang and X.~H.~Ge,
``Black holes in 4D Einstein-Maxwell-Gauss-Bonnet gravity coupled with scalar fields,''
[arXiv:2011.08604 [hep-th]].

\bibitem{1205.0559}
D.~Kubiznak and R.~B.~Mann,
``P-V criticality of charged AdS black holes,''
JHEP \textbf{07}, 033 (2012)
doi:10.1007/JHEP07(2012)033
[arXiv:1205.0559 [hep-th]].


\bibitem{1702.06766}
R.~G.~Cai, M.~Sasaki and S.~J.~Wang,
``Action growth of charged black holes with a single horizon,''
Phys. Rev. D \textbf{95}, no.12, 124002 (2017)
doi:10.1103/PhysRevD.95.124002
[arXiv:1702.06766 [gr-qc]].


\bibitem{1208.6251}
S.~Gunasekaran, R.~B.~Mann and D.~Kubiznak,
``Extended phase space thermodynamics for charged and rotating black holes and Born-Infeld vacuum polarization,''
JHEP \textbf{11}, 110 (2012)
doi:10.1007/JHEP11(2012)110
[arXiv:1208.6251 [hep-th]].



\end{thebibliography}
\end{document}